\documentclass[prl,reprint,amssymb,amsmath,aps,superscriptaddress,showpacs,showkeys]{revtex4}
\usepackage[linkcolor=blue]{hyperref}
\usepackage{graphicx}
\usepackage{color}
\usepackage{url}
\usepackage{amsmath}

\newcommand{\be}{\begin{equation}}
\newcommand{\ee}{\end{equation}}
\newcommand{\dif}{\mathrm{d}}
\newcommand{\rv}{\vec{r}}
\newcommand{\nv}{\vec{n}}
\newcommand{\Fv}{\vec{F}}
\newcommand{\vv}{\vec{v}}
\newcommand{\R}{\mathcal{R}}
\newcommand{\G}{\mathcal{G}}

\begin{document}


\title{ Anomalous segregation dynamics of self-propelled particles }
\author{ Enys Mones }
\email{ enys@hal.elte.hu }
\affiliation{ Department of Biological Physics, E\"{o}tv\"{o}s Lor\'{a}nd University, P\'{a}zm\'{a}ny P\'{e}ter stny. 1/A, H-1117 Budapest, Hungary }
\author{ Andr\'{a}s Czir\'{o}k }
\email{ andras@biol-phys.elte.hu }
\affiliation{ Department of Biological Physics, E\"{o}tv\"{o}s Lor\'{a}nd University, P\'{a}zm\'{a}ny P\'{e}ter stny. 1/A, H-1117 Budapest, Hungary }
\affiliation{ Department of Anatomy \& Cell Biology, University of Kansas Medical center, Kansas City, KS, USA }
\author{ Tam\'{a}s Vicsek }
\email{ vicsek@hal.elte.hu }
\affiliation{ Department of Biological Physics, E\"{o}tv\"{o}s Lor\'{a}nd University, P\'{a}zm\'{a}ny P\'{e}ter stny. 1/A, H-1117 Budapest, Hungary }
\affiliation{ Biological Physics Research Group of HAS, P\'{a}zm\'{a}ny P\'{e}ter stny. 1/A, H-1117 Budapest, Hungary }
\date{\today}

\begin{abstract}
A number of novel experimental and theoretical results have recently been obtained on active soft matter, demonstrating the various interesting universal and anomalous features of this kind of driven systems. Here we consider a fundamental but still unexplored aspect of the patterns arising in the system of actively moving units, i.e., their segregation taking place when two kinds of them with different adhesive properties are present. The process of segregation is studied by a model made of self-propelled particles such that the particles have a tendency to adhere only to those which are of the same kind. The calculations corresponding to the related differential equations can be made in parallel, thus a powerful GPU card allows large scale simulations. We find that in a very large system of particles, interacting without explicit alignment rule, three basic segregation regimes seem to exist as a function of time: i) at the beginning the time dependence of the correlation length is analogous to that predicted by the Cahn-Hillard theory, ii) next rapid segregation occurs characterized with a separation of the different kinds of units being faster than any previously suggested speed, finally, iii) the growth of the characteristic sizes in the system slows down due to a new regime in which self-confined, rotating, splitting and re-joining clusters appear. Our results can explain recent observations of segregating tissue cells {\it in vitro}.
\end{abstract}

\maketitle

\def\noteq{\neq}

\section{I. INTRODUCTION}

Collective flow of self-propelled biological units is observed on many scales ranging from molecular motor-driven cytoskeletal polymers, microscopic organisms, tissue cells and animals \cite{VicsekZafeiris12}. Non-living objects under specially designed conditions, such as fluidized granular materials \cite{Kudrolli08}, surface tension difference-driven ``Janus particles'' \cite{Palacci10b,Buttinoni13}, light pushed colloids \cite{Ni13} or micron-sized edges \cite{Buzas12} or spheres rotating due to an applied electric field \cite{Bricard13} can also behave as self-propelled particles. All these systems, active fluids, exhibit unusual behavior like the emergence of long-range correlations, superdiffusive behavior, anomalous density fluctuations, or explosive aggregation dynamics \cite{Marchetti13}. 

Here we focus on the unusual segregation dynamics of a mixture of different kinds of self-propelled particles at high density surface coverage. Segregation may occur due to the difference in the motilities of the particles \cite{Kabla12,McCandlish12} 
but in inanimate systems it is typically driven by adhesion differences, like the preferred adhesion to objects of the same type, and manifests as a slow coarsening described by the Cahn-Hilliard equation \cite{Bray02}.  In binary mixtures the characteristic domain size $\lambda$ grows in time, $t$, as 
 \be
 \lambda \sim t^z
 \ee
with $z=1/3$ for even coverage ratios, and $z=1/4$ for unequal coverage rations \cite{Nakajima11}. Yet, this dynamics is expected to change when the particles have an intrinsic motility, which is not driven by a free energy gradient of the whole system. For example, ballistically moving clusters are expected to aggregate faster \cite{Cremer14}. Yet the presence of slower moving objects, and the corresponding excluded volume constraints can make this process different from the aggregation of self-propelled clusters from a low density phase. In addition, the slower objects may also organize into a confinement barrier, strongly influencing the flow of more active particles \cite{Kabla12}.

Such segregation settings are also achievable in most experiments using non-living self-propelled objects. The interaction between highly active and more stationary (but not necessarily less adhesive) units is also relevant in crowd control \cite{Silverberg13}, but perhaps most obviously in multicellular systems. Cells of the same type segregate into disjunct clusters during various stages of embryonic development, often utilizing both adhesion and motility differences \cite{MehesVicsekReview13}. 

To explain this phenomenon, the differential adhesion hypothesis (DAH) \cite{Steinberg63} proposed that cell types have distinct adhesion properties. However, while the DAH and the corresponding computational models yield a slow coarsening similar to Cahn-Hilliard exponents \cite{Nakajima11}, recent experiments revealed a much faster segregation dynamics, with 
 \be
 1/2 < z < 1
 \label{zexp}
 \ee
for both in 3D spheroid \cite{Beysens00} and 2D monolayer \cite{Mehes12} cultures. This increased coarsening exponent is likely to contribute to the surprisingly fast development of early embryos.

Anomalous sorting was reported within a self-propelled particle system in which particles actively align their motion directions and interact with a hard core short range repulsion and a somewhat longer, but still short range attraction \cite{Belmonte08}. This pioneering study indicated, for the first time, that ​taking into account the self-propelled nature of the units leads to qualitatively new behavior as compared to simple diffusion-like motion. While the exponent $z$ was not determined directly, an analogous measure indicates $z\sim 0.18 < 1/3$ as discussed in \cite{Nakajima11}. Here we demonstrate that the self-propelled nature of the units (i.e., active cell motility) can substantially speed up the Cahn-Hilliard segregation dynamics and can yield behavior compatible with the experimentally observed $z$ values in 2D cell cultures (\ref{zexp}).  Following \cite{Szabo06}, we define a self-propelled particle model of binary mixtures. In the considered model particles interact with short range inter-particle forces and also adjust the direction of self-propulsion in response to these forces. By large scale computer simulations we explore the segregation dynamics and the spontaneously developing velocity fields. Our simulations indicate that in a very large system of particles, interacting without an explicit alignment rule, three basic segregation regimes seem to exist as a function of time: i) at the beginning the time dependence of the correlation length is analogous to that predicted by the Cahn-Hillard theory, ii) next rapid segregation occurs characterized with a separation of the different kinds of units being faster than any previously suggested speed, finally, iii) the growth of the characteristic sizes in the system slows down due to a new regime in which self-confined, rotating, splitting and re-joining clusters appear.

\section{II. THE SEGREGATION MODEL}

Self-propelled particles (SPPs) are modeled using the following two-dimensional overdamped equation of motion:
 \be
 \frac{\dif\rv_i}{\dif t}=v_0\nv_i+\Fv_i,
 \label{eom}
 \ee
where $\rv_i$ is the position of the $i$th particle, $v_0$ and $\nv$ are the magnitude and direction of its active motion and $\Fv$ denotes the (net) force acting on the particle resulted by short range interactions with its environment.
Forces are exerted by other particles and the boundary, modeled as a repulsive rigid wall:
 \be
  \Fv_i=\sum_{j=1}^N\Fv^\mathrm{cell}(\rv_i,\rv_j)+\Fv^\mathrm{wall}(\rv_i).
 \label{forces}
 \ee

Following \cite{Szabo06}, we model the interparticle force with a piecewise linear function of the distance $d_{ij}=|\rv_i-\rv_j|$. This force represents the combination of a shorter range repulsive and longer range adhesive interaction as
 \be
  \Fv^\mathrm{cell}(\rv_i,\rv_j)={\rv_j-\rv_i\over d_{ij}}\left\{
  \begin{array}{l l}
    F_\mathrm{rep}\frac{d_{ij}-D_\mathrm{eq}}{D_\mathrm{eq}}, & \text{if $d_{ij}<D_\mathrm{eq}$,} \\
    F_\mathrm{adh}\frac{d_{ij}-D_\mathrm{eq}}{R_0-D_\mathrm{eq}}, & \text{if $D_\mathrm{eq}<d_{ij}<R_0$,}\\
    0, & \text{otherwise}.
  \end{array} \right.
 \label{force}
 \ee
In expression (\ref{force}) the interaction range is $R_0$, and $D_\mathrm{eq}$ denotes the equilibrium distance where the attractive and repulsive forces are balancing each other.
In the simulations, $R_0$ and $D_\mathrm{eq}$ are set so that the cells behave as soft disks and the interaction range is slightly larger than their diameter.
The coefficients $F_\mathrm{rep}$ and $F_\mathrm{adh}$ set the strength of the repulsion and adhesion, respectively.
We assumed that adhesion is absent between cells of different types ($F_\mathrm{adh}^{\R-\G}=0$), and the repulsion coefficient is fixed for all cell interactions.

The boundary exerts a force that is exponentially decreasing with the distance from the wall $d_{iw}$:
 \be
  \Fv^\mathrm{wall}(d_{iw})=\vec{e}_{iw}\left\{
  \begin{array}{l l}
    F_w\exp\Big(-\frac{2d_{iw}}{R_0}\Big), & \text{if $d_{iw}<R_0$,}\\
    0, & \text{otherwise},
  \end{array} \right.
 \ee
where $\vec{e}_{iw}$ is the unit vector pointing to the closest boundary segment and the $F_w$ coefficient sets the magnitude of this force.

Finally, SPP heading vectors are steered towards the particles' physical displacements by the rule proposed in \cite{Szabo06}:
 \be
 \frac{\dif\theta_i}{\dif t}= \mathrm{arg}(\vv_i)-\theta_i + \xi
 \label{steer}
 \ee
where $\theta_i=\mathrm{arg}(\nv_i)$ is the angle of heading vector $\nv_i$ and $\mathrm{arg}(\vv_i)$ is the angle of the actual velocity as influenced by interparticle forces. This steering is, however,  imperfect -- which is modeled by a Gaussian white noise with zero mean and finite variance 
 \be
 \langle\xi(t)\xi(t')\rangle={\eta^2\over 12}\delta(t,t').
 \ee
The motivation for such a steering rule comes from observations indicating that cells can respond to mechanical forces \cite{Verkhovsky99, Yam07, Kozlov07, Tambe11, Weber12}. It is also consistent with a positive feedback regulation of front-rear cell polarity by actual cell displacements (actin polymerization in the front of the cell) \cite{Dawes07} as discussed further in \cite{Szabo10}.

We compare the above defined SPP model with a similar conventional Brownian particle model in which particles move in completely random directions:
 \be
  \nv_i(t)=\vec{\xi},
 \label{ndp}
 \ee
where the noise $\vec{\xi}$ is a two-dimensional random vector with zero mean and finite variance.

To demonstrate the importance of the steering rule (\ref{steer}), we also performed simulations with persistent Brownian particles, where Eq.~\ref{steer} has been replaced by
 \be
 \frac{\dif\theta_i}{\dif t}= \xi.
 \label{nosteer}
 \ee

\section{III. PARAMETERS}

Simulations were carried out in a closed rectangular domain of size $L$ with repulsive walls.  We investigated the segregation of two different types of particles -- being referred to below as red ($\R$) and green ($\G$). Model parameter values were assigned to particles depending on which group they belonged to. The choices of the parameter values for the two particle types were motivated by the observations of cell sorting in monolayers \cite{Mehes12}. Thus, the red particles are larger, more motile and have stronger adhesive force among each other. We assumed that while short-range repulsion exists between each particle, the attractive force acts only between particles of the same type.  

The number of particles was set to obtain a close to fully packed coverage of the entire simulation area (mostly in the range of 90\% coverage). While the average particle ``size'' ($D_\mathrm{eq}$) was determined by the particle type ($D_\mathrm{eq}=1$ for red and $D_\mathrm{eq}=0.5$ for green particles), the actual size of each particle was randomly selected  from a uniform distribution between $[D_\mathrm{eq}-0.1, D_\mathrm{eq}+0.1]$. Thus both particle groups were polydisperse. In a polydisperse system the interaction (\ref{force}) is evaluated using the average $R_{eq}$ values of the interacting partners. The interaction range was set as $R_0=1.4 D_\mathrm{eq}$. The number of particles of a given type was determined from a target surface coverage. As an example, to achieve a 1:1 coverage ratio, the number of particles was inversely proportional to their size: ${N_\R}/{N_\G}={D^2_\G}/{D^2_\R}$.

The natural length scale of the simulations is the mean diameter of a (red) particle, $D_\mathrm{eq}=1$. The natural time unit of the simulations, $\tau$, is the time that is needed for a cell to move a distance of the average diameter. Since this time depends on a number of model parameters, our time units are fixed as the natural time units of the simulations corresponding to the parameter values given in Table 1.

As reported previously \cite{Szabo06}, the main control parameter of the model is the precision of the steering term. When adaptation of the heading and movement vectors is precise  (i.e., the noise in Eq.~(\ref{steer}) is small), the system organizes into a long-range ordered state. In this state particle velocities are correlated over distances that are comparable with the system size. As the simulations are carried out in a finite domain with repulsive walls, in the long-range ordered regime the whole system rotates in a randomly selected direction. Since this state was not directly observed in cell sorting experiments, for our simulations we have chosen a parameter setting which does not lead to correlated motion within the entire simulation domain, yet the steering is precise enough to build velocity correlations over substantial distances.

 \begin{table*}
 \begin{center}
 \begin{tabular}{lp{1cm}lp{1cm}r}
 Parameter description                 && Name                             && Value \\
 \hline
 red cell velocity       && $v_\R$                           && 3.125 \\[3pt]
 green cell velocity     && $v_\G$                           && 1.25 \\[3pt]
 repulsive force coeff.                && $F_\mathrm{rep}$                        && 187.5 \\[3pt]
 red-red adhesive force coeff.         && $F_\mathrm{adh}^{\R-\R}$                && 30 \\[3pt]
 green-green adhesive force coeff.     && $F_\mathrm{adh}^{\G-\G}$                && 7.5 \\[3pt]
 noise                && $\eta_0\sqrt{\Delta t}$          && $7^\circ$ in degrees\\[3pt]
 \hline
 \end{tabular}
 \end{center}
 \caption{ The set of parameters appearing in Eqs.~(\ref{eom})-(\ref{steer}), along with their corresponding symbols and simulational values. }
 \end{table*}

\section{IV. RESULTS}
\subsection{A. Dynamic exponents from simulations}

\begin{figure}
\begin{center}
\includegraphics[width=5in]{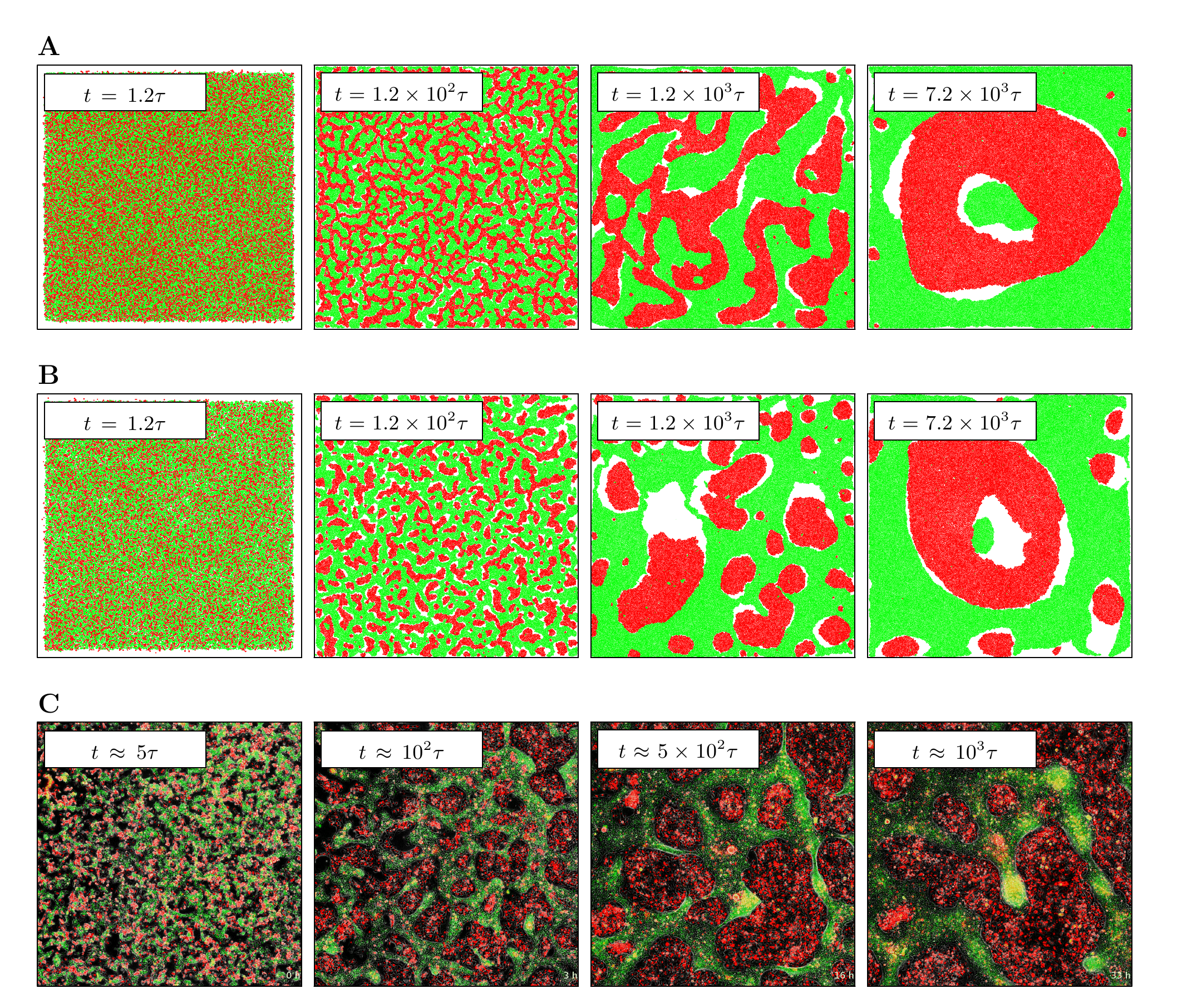}
\caption{\small
Morphologies characterizing the segregation of a SPP mixture at 50:50 (a) and 40:60 (b) coverage ratios. Red particles are more motile than green particles (see Table I, $N=10^5$, $L=L_0=100$).
White areas are devoid of particles -- uniform clusters can achieve higher local cell density than areas where the two particle types are intermixed and their movement is less correlated. In the final state of the simulation the red cluster rotates (see supplemental material). As a comparison, we show characteristic images from the experiment of \cite{Mehes12} (c). 
Time unit $\tau$ corresponds to the time needed for an SPP particle or cell to move a distance that is equal to the average particle/cell size. 
}
\label{fig1}
\end{center}
\end{figure}

\begin{figure}
\begin{center}
\includegraphics[width=5in]{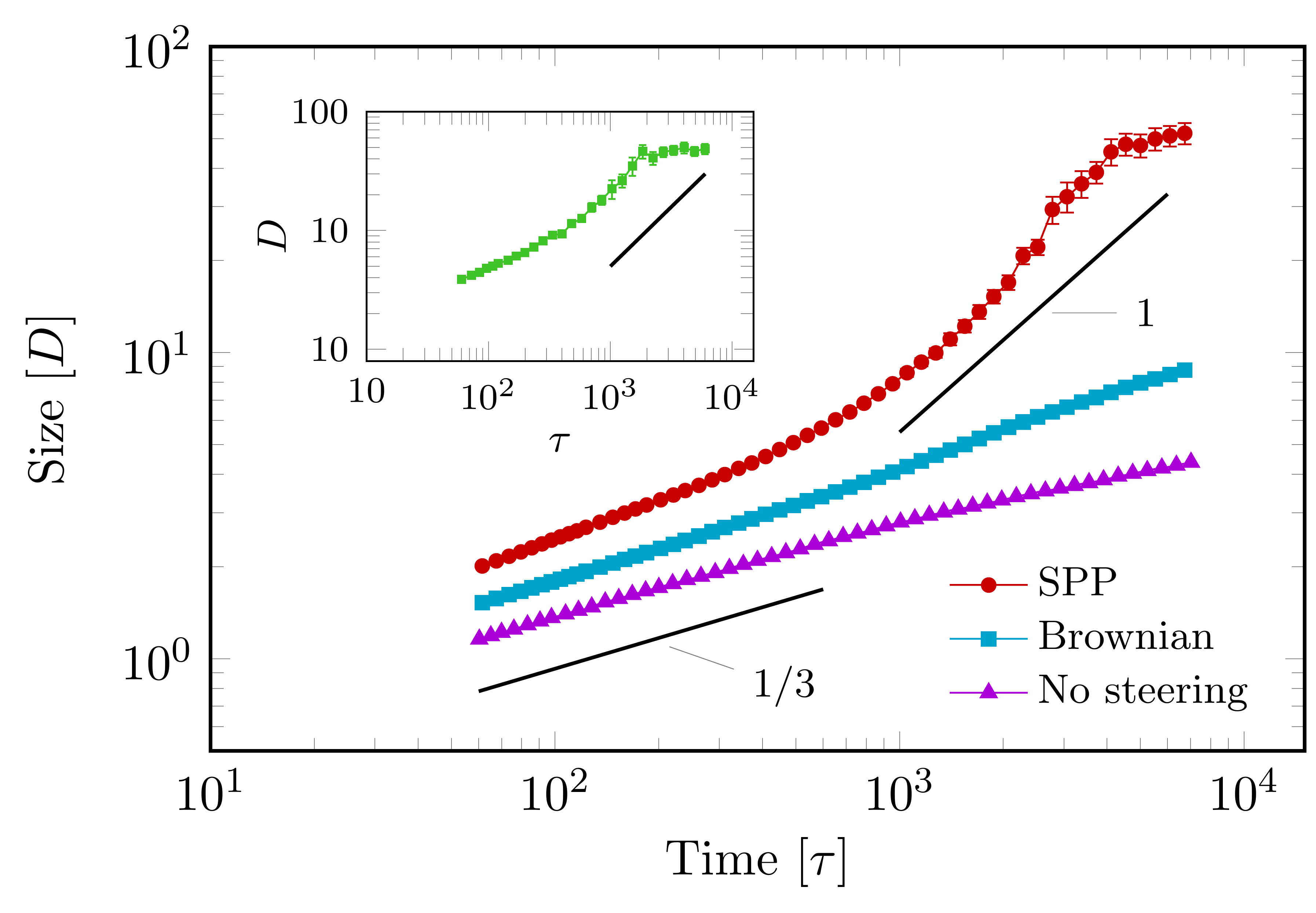}
\caption{\small
An SPP system can segregate much faster than a similar system containing noise-driven (Brownian) particles. In Brownian simulations the characteristic linear size of the segregated domains grows according to the Cahn-Hilliard behavior. In contrast, the SPP system exhibits a regime where the average cluster size is proportional to time. Fast segregation is also observed in simulations where all particles have identical properties (values characterizing red particles in Table I), and the segregation is driven only by the lack of adhesion between red and green particles (inset, noise=$6.3^\circ$, coverage: 70\%). Persistent random motility without the specific steering rule (\ref{steer}) also exhibits Cahn-Hilliard coarsening. The solid lines are guides to the eye. Spatial scale unit is the mean particle diameter, temporal unit is the time an SPP needs to move a unit distance. Error bars represent standard error of the mean ($\leq n \leq 12$).
}
\label{fig2}
\end{center}
\end{figure}

\begin{figure}
\begin{center}
\includegraphics[width=5in]{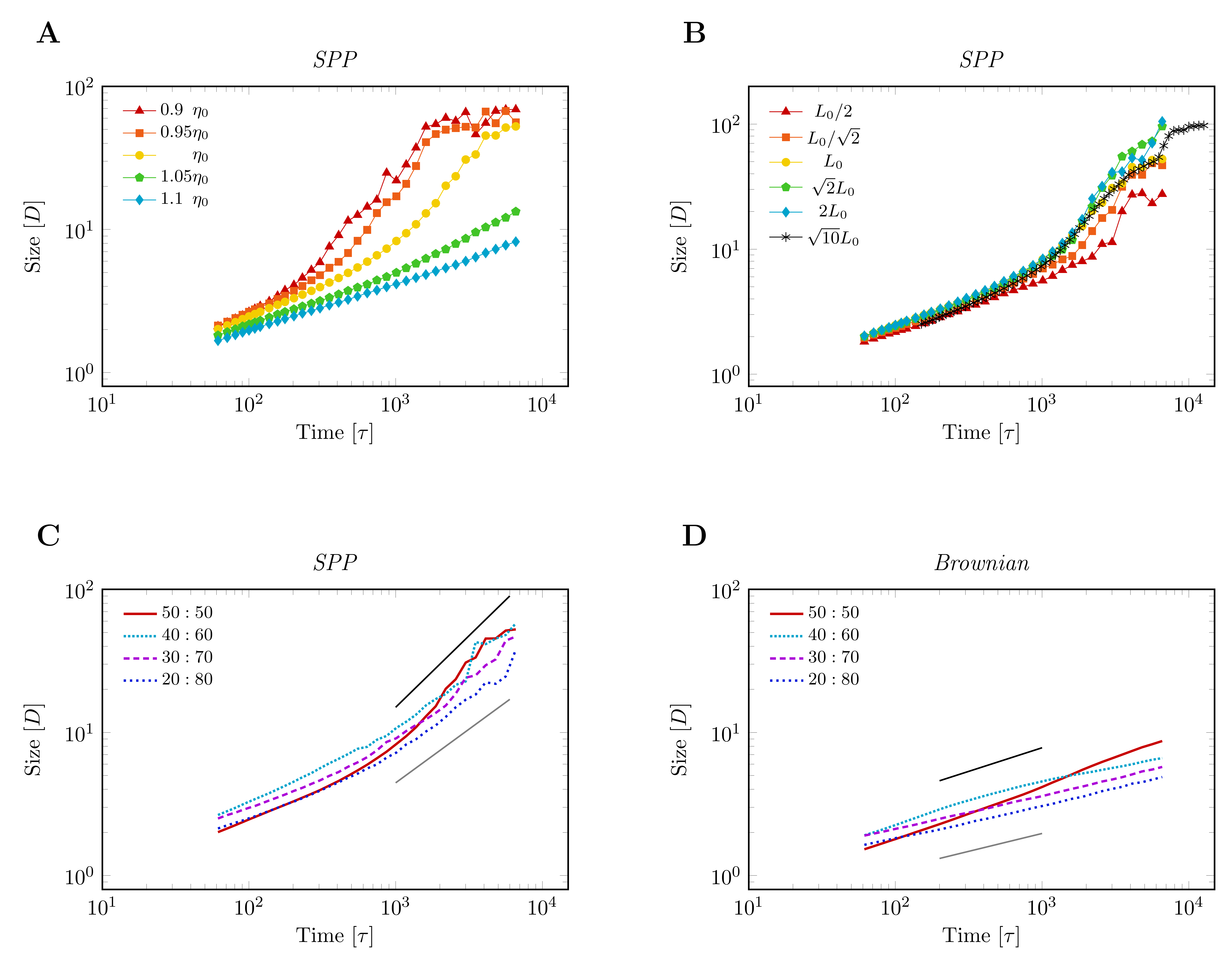}
\caption{\small
Cluster size dynamics for various model parameter values. a: Better steering quality (decreased noise) yields earlier and faster segregation. The transition between the fast ($z\approx1$) and slow ($z\approx1/3$) mechanism is sudden (elicited by a 10\% change in the noise parameter) and is coincident with the transition between a long-range ordered (rotating) and a locally ordered, but globally disordered system. In the transient regime (red symbols) the velocity correlation length is still smaller than the system size, yet the segregation is much faster than the Cahn-Hilliard behavior. B: Maximal cluster size is limited by the system size. For larger systems the linear growth regime is extended. C: When the coverage ratio differs from 1:1, the segregation is slower than the linear growth shown in panels A and B, yet it is still faster than the Cahn-Hilliard behavior. D: As a comparison, noise driven particle system exhibit Cahn-Hilliard segregation with $z\approx1/3$ for 1:1 coverage ratio and $z\approx1/4$ otherwise. 
}
\label{fig3}
\end{center}
\end{figure}

\begin{figure}
\begin{center}
\includegraphics[width=5in]{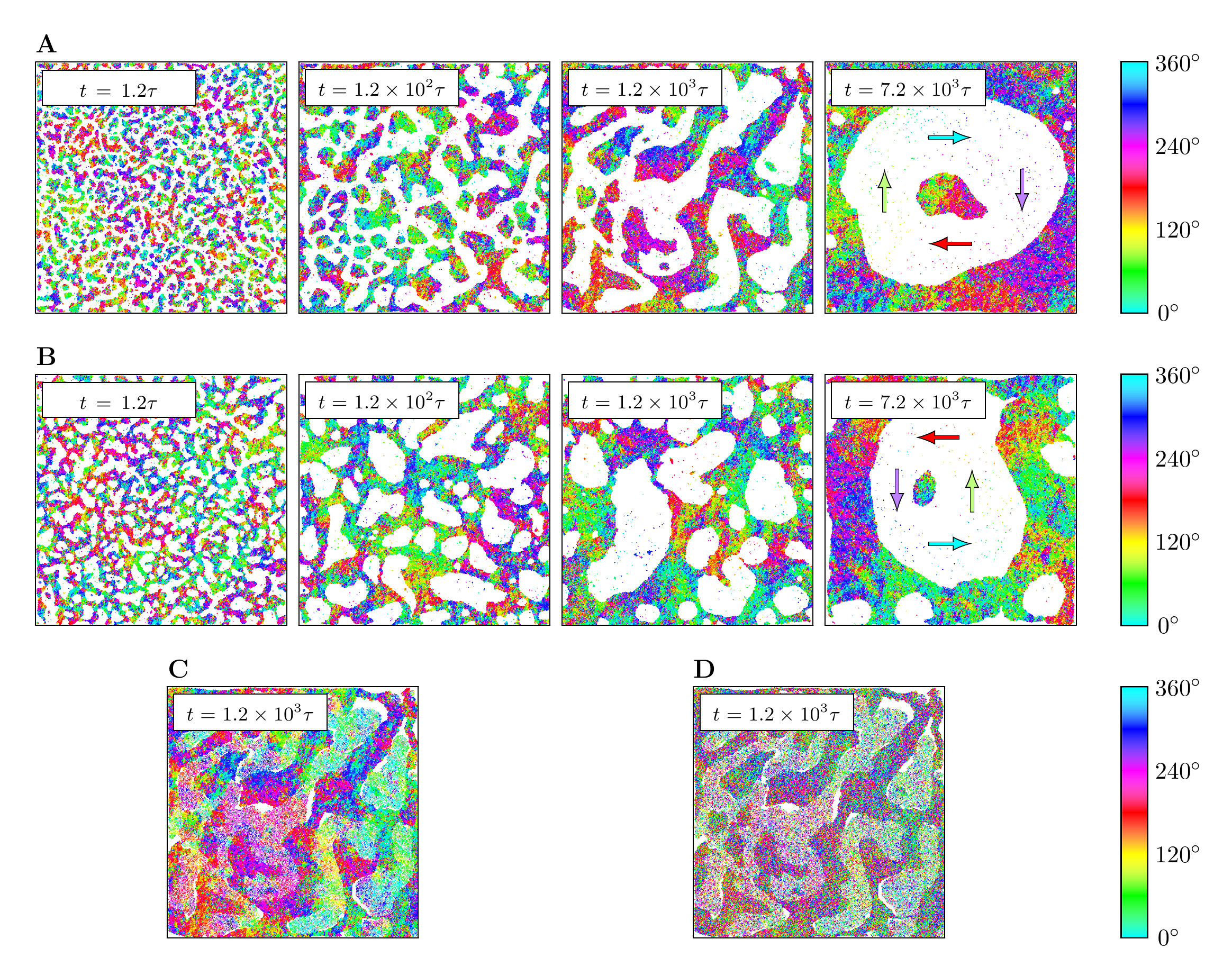}
\caption{\small
Directions of particle movements at various stages of the segregation process.  Panels (A) and (B) depict color coded velocity directions of the green particles shown in Fig 1, hence SPP mixtures at 50:50 (A) and 40:60 (B) coverage ratios. C: Velocity directions of both red and green particles within the system shown in panel A. The abrupt change of motion direction at segregation boundaries indicates that segregated cell groups can slide against each other. D: Heading directions in the same snapshot shown in panel C. Gray colors indicate that heading directions are less correlated locally than actual displacements are.
}
\label{fig4}
\end{center}
\end{figure}

Simulations started with a random binary mixture readily segregate irrespective of the coverage ratio (Fig.~\ref{fig1}). As the supplemental movies demonstrate, clusters of segregated particles are highly motile and may even reach a state where all particles segregate into a single, rotating cluster. Snapshots of cell cultures in Fig.~\ref{fig1} show the 2D segregation  of EPC (green) and primary fish keratocytes (red) obtained from \cite{Mehes12}. To derive the natural time unit $\tau$ for the experiments, we considered that the diameter of fish keratocytes is $D\approx20\mu$m, and have a persistent velocity of $v_0\approx500\mu$m/h. Thus, the characteristic time -- needed for a displacement the size of a unit cell diameter -- in the experiment is $\tau\approx2\mathrm{min}$. When time is measured in the natural unit, the corresponding simulated and experimentally observed configurations are very similar. 

The characteristic spatial scale $\lambda$ of a given configuration  was determined by wavelet transform. Segmented images of the simulated system were convolved with a series of two-dimensional Mexican hat wavelets -- each with a distinct scale parameter. The characteristic scale was set as the wavelet parameter that yields the largest standard deviation in the convolved image. Repeating this procedure for a series of configurations was used to characterize the time development of the segregation process. As Fig.~\ref{fig2} demonstrates for an even coverage of the two particle types, both the Brownian and the persistent/ballistic (\ref{nosteer}) systems exhibit the characteristic Cahn-Hilliard behavior with $z=1/3$. In contrast, the segregation of the SPP simulation is triphasic. After an initial, Cahn-Hilliard-like behavior a fast segregating regime sets in with $z\approx 1$. In this regime the particles rapidly organize into giant clusters. Finally, sufficiently large  simulations reach a new regime characterized by self-confined, rotating clusters that may split and re-join depending on their internal dynamics. While particles stream within these clusters, the whole system -- composed of the two kinds of particles -- do not exhibit long range-ordered movements.

The same behavior also emerges in simulations with much simpler parameter choices, when the segregation is driven only by the lack of adhesion between the red and green particles (Fig.~2 inset). However, most of our simulations were carried out for parameter values  intended to correspond to the conditions of prior experiments on cell sorting in order to make a comparison possible. To investigate how model and simulation parameters affect the segregation, the system size, noise and the red-green coverage ratio was systematically varied from the parameter setup used to generate Figs 1-2.  As Fig.~\ref{fig3}A demonstrates, the segregation is faster and its onset is earlier when steering noise is decreased. The transition between the fast ($z\approx1$) and slow ($z\approx1/3$) regimes is sudden: it is elicited by a 10\% change in the noise amplitude. The transition between coarsening regimes is coincident with the transition between a long-range ordered (rotating, blue and orange symbols in Fig.~\ref{fig3}A) and a locally ordered, but globally disordered system (yellow and green symbols in Fig.~\ref{fig3}A). In a transient regime (red symbols in Fig.~\ref{fig3}A) the velocity correlation length is still smaller than the system size, yet the segregation is much faster than the Cahn-Hilliard behavior. 

Simulations performed with various system sizes (Fig.~\ref{fig3}B) reveal an intrinsic (system size independent) upper limit for the linear cluster growth regime.  For uneven coverage ratios the coarsening is slower both in the SPP (Fig.~\ref{fig3}C) and the Cahn-Hilliard-type Brownian system (Fig.~\ref{fig3}D). Still, coarsening in the  the SPP system continues to be fast as the table of exponents indicate.

  \begin{table}
  \begin{center}
  \begin{tabular}{lp{1cm}r}
  Dynamics      && Exponent \\
  \hline
  SPP (50:50)   && 1.14 $\pm$ 0.12 \\
  SPP (uneven)  && 0.75 $\pm$ 0.05 \\
  Brownian (50:50)   && 0.38 $\pm$ 0.02 \\
  Brownian (uneven)  && 0.24 $\pm$ 0.02 \\
  \hline
  \end{tabular}
  \end{center}
  \caption{ Power-law exponents in the different models (SPP and Brownian), with even and uneven volume fractions.
  The values correspond to the averages of fits to 4 trends, where each trend is averaged over 3 independent measurement.
  Errors idicate the standard deviations.}
  \end{table}
 
The results shown in Fig.~\ref{fig3} thus indicate that fast segregation is characteristic for the locally ordered, but globally disordered regime of the SPP system.
As spatially correlated movements seem to be of key importance, in Fig.~\ref{fig4} we show velocity field snapshots that are representative for various stages of the segregation process. The parallel increase of the velocity correlation length and pattern size indicate that in the SPP model clusters can move coherently for extended distances. In contrast, a local diffusive movement is characteristic for the Cahn-Hilliard-type Brownian system. The velocity snapshots also reveal that clusters can readily glide along each other as motion directions can change abruptly at cluster boundaries (Fig.~\ref{fig4}C).

\subsection{B. Qualitative Interpretation}

The different exponents in the Brownian and SPP system can be qualitatively understood if we compare the predominantly diffusive (Brownian) and ballistic (SPP) aggregation kinetics \cite{Stenhammar13}. Let us consider the system at time $t_1$ when the characteristic size of the motile clusters is $\lambda_1$. By time $t_2$ the typical cluster size is increased to $\lambda_2=\sqrt{2}\lambda_1$  reflecting the coalescence of two clusters of size $\lambda_1$. If the typical distance separating the clusters was $\lambda'_1$ at time $t_1$, it is increased to $\lambda'_2=\sqrt{2}\lambda'_1$ at time $t_2$ to keep the average particle density $\lambda^2/(\lambda+\lambda')^2$ constant. Hence, during the coarsening we expect the size of the clusters scale with the distance separating the clusters: $\lambda\sim\lambda'$. 

During a sufficiently small time interval $\Delta t$, each particle within a noise driven cluster is subjected to a random displacement, $v_0\int_0^{\Delta t}\vec{\xi}(\tau)d\tau$ according to Eqs (\ref{eom}) and (\ref{ndp}). For a cluster of $N\sim\lambda^2$ particles the vectorial sum of these displacements is a random vector with zero mean and standard deviation $v_0\eta\sqrt{N\Delta t}$. If the attractive forces within the cluster are strong enough to maintain its integrity, the cluster's center of mass performs a random walk: the standard deviation of the center of mass displacement is $v_0\eta\sqrt{\Delta t/N}$. After an elapsed time $t$ (thus after $t/\Delta t$ steps) the variance of the center of mass is $v_0^2\eta^2t/N$  . Hence, the cluster's diffusion coefficient depends on its size $\lambda$ as
 \be
 D(\lambda)={v_0^2\eta^2\over 4N}\sim\lambda^{-2}.
 \ee
The time expected to move the cluster over a distance $\lambda'_1$ is
 \be
 t_2-t_1 \sim {{\lambda'_1}^2\over D(\lambda_1)} \sim \lambda^4_1.
 \ee 
As $t_2-t_1\sim\lambda_1^{1/z}$, for the aggregation of Brownian clusters we obtain the Cahn-Hilliard result of $z=1/4$.

In contrast, particles in the SPP model move superdiffusively as for their displacement $d$ over time $t$ 
 \be
d \sim t^{\alpha}
 \ee
holds with $\alpha\approx1$. Due to the extended spatial range of velocity correlations the cluster speed only weakly depends on cluster size. Thus
 \be
 \lambda'_1 \sim (t_2-t_1)^\alpha
 \ee
yielding $z=\alpha$. For ballistic motion $\alpha=1$, in our simulations $\alpha\approx 0.8$. 

The above arguments assumed that the segregation process is driven by the merger of individual clusters. When the coverage ratio of the two particle types is 50:50, the system organizes into stripes instead of clusters. In the Brownian case the value of the dynamic exponent $z=1/3$ corresponds to a situation in which the coarsening is driven by the lateral movement of the stripes performing a random walk with a diffusivity
$ D(\lambda) \sim\lambda^{-1}$.

\section{V. DISCUSSION}

Active motion within a confluent (densely packed) system is susceptible to jamming \cite{Henkes11}. While the polydisperse particle size distribution helps to prevent the formation of a crystalline lattice, the formation of coherently moving clusters for low noise are similar to the ``glider'' structures that develop in high density lattice gas simulations \cite{Peruani11}. An off lattice, overdamped SPP model with excluded volume interaction was recently shown to undergo spontaneous phase separation, by which particles create a dense cluster that is immersed in a ``gas phase'' of low particle density. Such a system, without the specific steering rule (\ref{steer}) linking propulsion direction to physical displacements, exhibits an aggregation kinetics with $z=1/2$ \cite{Redner12}. Recent scaling analysis of the irreversible SPP coagulation process indicated the possibility for $z>1$ and in certain cases even explosive cluster growth where the cluster size increases exponentially in time \cite{Cremer14}. While the high particle density and the reversible association of the particles prevents the realization of such explosive coarsening in our simulations, these recent result indicate that self-propelled systems can coarsen substantially faster than the Cahn-Hilliard behavior.

A comparison of our results to that of \cite{Belmonte08} and \cite{Fily12} reveals that the actual steering mechanism of the particles is just as important as the fact that they are self-propelled. This difference is partially captured by models operating with polar or apolar alignment rules: in the former  particles tend to move in a parallel direction, while in the latter particles can readily glide in opposite directions, like self-propelled rods \cite{Peruani06,Ginelli10,Peruani12}. Here there is no explicit ``polar'' coupling between the velocity of adjacent particles, yet they readily organize into a polar order. Yet, the segregation behavior is rich and surprisingly dependent on various model properties. For example, the simplest, symmetrical choice of parameteres (same adhesion between red and green particles, same motility) segregate slowly due to jamming. In contrast, the same parameter settings can yield $z\approx 1$ when the coverage is slightly decreased (Fig.~2, inset). Thus, fast segregation depends on the emergence of fast moving clusters. When the persistent motion of clusters is hindered either by jamming, extensive mixing, or the lack of cohesion -- the slower, diffusive Cahn-Hilliard behavior appears.

Embryonic morphogenesis and tissue regeneration are fascinating, complex processes. Cell sorting is one of the fundamental concepts that help us understand how multicellular patterns form at various stages of development. During cell sorting cells of the same type segregate into disjunct clusters. To explain this phenomenon, the differential adhesion hypothesis (DAH) \cite{Steinberg63} proposed that cell types can have distinct adhesion properties. Subsequent experiments demonstrated that a surface tension-like quantity can be assigned to cohesive cell clusters and this quantity predicts the spatial arrangement of cell sorting experiments \cite{Foty96}. The biophysical mechanism behind the macroscopic surface tension can be traced back to the amount of adhesion molecules, primarily cadherins, on the cell surface \cite{Foty05,Hegedus06}, and a spatially restricted force generation within the cortical cytoskeleton \cite{Maitre12}.  The clarity of the DAH made it well suited for theoretical models which assumed that changes in cell configuration are driven by minimization of a quantity analogous to surface energy. This insight led to the widespread use of the Cellular Potts Model (CPM) to describe dynamics of multicellular systems \cite{Glazier93}, or a variety of multi-particle models which represent intercellular adhesion by a combination of short range repulsive and a somewhat longer range attractive forces \cite{Newman05,Szabo06,Szabo07}. 

While the fast ($z \approx 1$) segregation of cells has been demonstrated using three dimensional aggregates \cite{Beysens00}, similarly fast kinetics was indicated in recent (two dimensional) monolayer cell culture  experiments as well \cite{Mehes12}. Although the fast segregation in 3D could be a consequence of hydrodynamic effects \cite{Siggia79} such as mechanical pressure created by the cells, the firm adhesion between the cells and the culture substrate in monolayer experiments \cite{Mehes12} points to the importance of active cell movements in the process. In this work we demonstrated using a previously proposed SPP model for multicellular behavior that suitable steering of active motility can indeed result in fast segregation with a kinetics compatible with experimental observations. The model is biologically plausible in the sense that the steering rule does not assume the cells' ability to calculate the local average direction of motion -- a common assumption in SPP models. Instead, the effect of adjacent cells is deduced indirectly through the (in)ability to move in certain directions.

\section{ACKNOWLEDGEMENT}
Funds from the European Union ERC COLLMOT Project, from the Hungarian Science Fund (K72664), the Hungarian Development Agency (KTIA AIK 12-1-2012-0020) and from the NIH (R01HL087136) are gratefully acknowledged.

\end{document}